\documentclass{emulateapj}

\newcommand{\kpch}{\>h^{-1} {\rm kpc}}

\def\gsim { \lower .75ex \hbox{$\sim$} \llap{\raise .27ex \hbox{$>$}} }
\def\lsim { \lower .75ex \hbox{$\sim$} \llap{\raise .27ex \hbox{$<$}} }

\shorttitle{Disk Galaxy Rotation Curves in Triaxial CDM Halos}
\shortauthors{Hayashi et al.}

\begin{document}

\title{Disk Galaxy Rotation Curves in Triaxial CDM Halos}

\author{ Eric Hayashi,\altaffilmark{1} Julio
 F. Navarro,\altaffilmark{1,2} Adrian Jenkins,\altaffilmark{3} Carlos
 S. Frenk,\altaffilmark{3} Chris Power,\altaffilmark{3} \\ Simon
 D.M. White,\altaffilmark{4} Volker Springel,\altaffilmark{4} Joachim
 Stadel,\altaffilmark{5} \\ Thomas Quinn,\altaffilmark{6} and James
 Wadsley\altaffilmark{7}}

\altaffiltext{1}{Department of Physics and Astronomy, University of Victoria, 
 Victoria, BC V8P 1A1, Canada}

\altaffiltext{2}{Fellow of CIAR and of the J.S.Guggenheim Memorial Foundation}

\altaffiltext{3}{Institute for Computational Cosmology, Department of Physics,
  University of Durham, South Road, Durham DH1 3LE, United Kingdom}

\altaffiltext{4}{Max Planck Institute for Astrophysics, Karl-Schwarzschild Strasse 1,
Garching, Munich, D-85740, Germany}

\altaffiltext{5}{Institute for Theoretical Physics, University of Zurich,
  Winterthurerstrasse 190, Zurich CH-8057, Switzerland}

\altaffiltext{6}{Department of Astronomy, University of Washington, Seattle, WA
98195, USA}

\altaffiltext{7}{Department of Physics and Astronomy, McMaster University, Hamilton,
ON L8S 4M1, Canada}

\begin{abstract}
We use N-body hydrodynamical simulations to study the structure of disks in
triaxial potentials resembling CDM halos. Our analysis focuses on the accuracy
of the dark mass distribution inferred from rotation curves derived from
simulated long-slit spectra.  We consider a massless disk embedded in a halo
with axis ratios of $0.5\colon0.6\colon1.0$ and with its rotation axis aligned
with the minor axis of the halo. Closed orbits for the gaseous particles deviate
from coplanar circular symmetry, resulting in a variety of long-slit rotation
curve shapes, depending on the orientation of the disk relative to the line of
sight. Rotation curves may thus differ significantly from the
spherically-averaged circular velocity profile of the dark matter
halo. ``Solid-body'' rotation curves---typically interpreted as a signature of a
constant density core in the dark matter distribution---are obtained about 25\%
of the time for random orientations although the dark matter follows the cuspy
density profile proposed by Navarro, Frenk \& White (NFW). We conclude that the
discrepancies reported between the shape of the rotation curve of low surface
brightness galaxies and the structure of CDM halos may be resolved once the
complex effects of halo triaxiality on the dynamics of the gas component is
properly taken into account. 
\end{abstract}


\section{Introduction}
\label{sec:intro}

It is commonly believed that the inner regions of low surface brightness (LSB)
galaxies are ideal probes of the inner structure of dark matter halos.  Given
the small contribution of the baryonic component to the mass budget in these
galaxies, dynamical tracers of the potential such as rotation curves are
expected to cleanly trace the dark matter distribution. This provides important
astrophysical clues to the nature of dark matter, since the spatial distribution
of dark material in these highly non-linear regions is expected to be quite
sensitive to the physical properties of the dark matter.

LSB rotation curves can thus be contrasted directly with theoretical predictions
of the inner structure of halos, and there is now an extensive body of work in
the literature that reports substantial disagreement between the shape of LSB
rotation curves and that of circular velocity curves of simulated cold dark
matter (CDM) halos \citep[see, e.g.,][]{FLORES94,MOORE94,MCGAUGH98,DEBLOK01}.
Some of these rotation curves are fit better by circular velocity curves arising
from density profiles with a constant density ``core'' rather than by the
``cuspy'' density profiles commonly used to fit the structure of CDM halos
(Navarro, Frenk \& White 1996, 1997, hereafter NFW). This discrepancy adds to a
growing list of concerns regarding the consistency of CDM with
observational constraints on the scale of individual galaxies \citep[see,
e.g.,][]{SELLWOOD00} that has prompted calls for a radical revision of the CDM
paradigm on small scales \citep[see, e.g.,][]{SS00}.

Before accepting the need for radical modifications to CDM it is important to
note a number of caveats that apply to the LSB rotation curve problem.  For
instance, many of the early rotation curves where the disagreement was noted
were significantly affected by beam smearing in the HI data \citep{SWATERS00}.
The observational situation has now improved substantially thanks to
higher-resolution rotation curves obtained from long-slit H$\alpha$ observations
\citep[see, e.g.,][]{DEBLOK02,SWATERS03}. We shall restrict our analysis to
these newer datasets in what follows.

We also note that, strictly speaking, the observational disagreement is with the
fitting formulae used to parameterize the structure of simulated CDM halos
(usually the profile proposed by NFW), rather than with the structure of
simulated halos themselves. Although the fitting formulae provide a simple and
reasonably accurate description of the mass profile of CDM halos, the radial
range over which they have been validated often does not coincide with the
scales where the disagreement has been identified.

Furthermore, small but significant deviations between the NFW profile and
simulated halos have been reported as the mass and spatial resolution of the
simulations has increased \citep{MOORE98,GHIGNA00,FUK97,FUK01}. Although there
is no broad consensus yet regarding how these deviations may affect the
comparison with observed rotation curves \citep[see,
e.g.,][]{POWER03,HAYASHI04,NAVARRO04}, the fact that the deviations worsen
towards the centre advise against using extrapolations of simple fitting
formulae such as the NFW profile to assess consistency with observation.

Finally, it must be emphasized that the ``cusp vs. core'' problem arises when
comparing rotation speeds of LSB disks to spherically-averaged circular
velocities of dark matter halos. Given that CDM halos are expected to be
significantly non-spherical
\citep{DAVIS85,FRENK88,JING95,JING02}, some differences
between the two are to be expected. It is therefore important to use the full 3D
structure of CDM halos to make predictions regarding the rotation curves of
gaseous disks that may be compared directly to observation.

\begin{figure}[t]
\epsscale{1.05}
\plotone{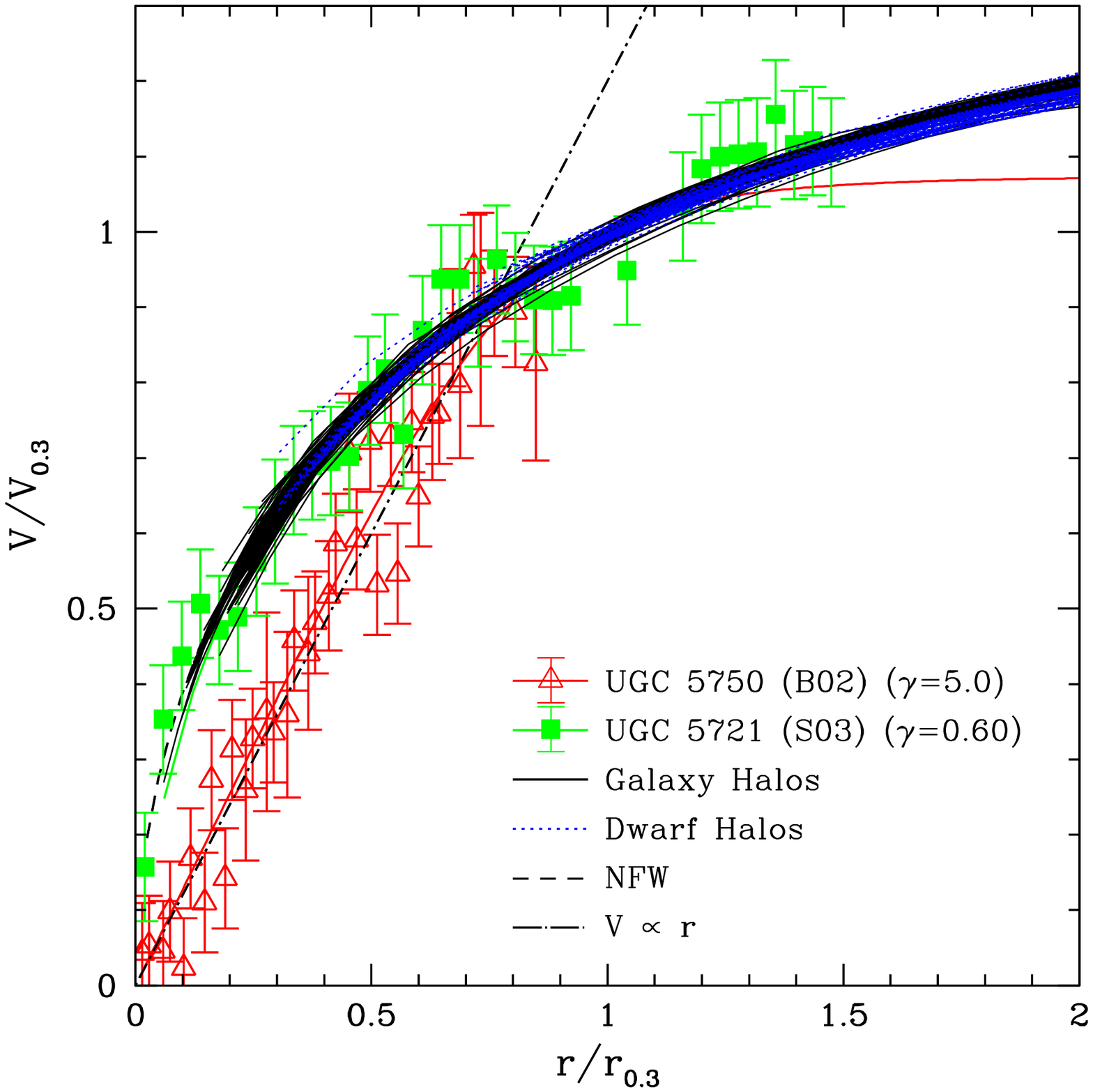}
\caption
[Rotation curves of LSB galaxies and circular velocity profiles of simulated halos]
{Rotation curves of two LSB galaxies from the samples of
    \citet[][B02]{DEBLOK02} and \citet[][S03]{SWATERS03}, chosen to illustrate
    their various shapes, as measured by the parameter $\gamma$
    from fits with the \cite{COURTEAU97} fitting formula.  Rotation curves have
    been scaled to the radius $r_{0.3}$ and corresponding velocity $V_{0.3}$
    (see text for details). Fits
    with large $\gamma$ values are characterized by a linear rise in velocity
    with radius followed by a sharp transition to the flat part of the curve.
    The NFW profile (dashed line) and the $V_c$ profiles of simulated dwarf-
    (dotted lines) and galaxy-sized halos (solid lines) match reasonably well
    systems with $\gamma \lsim 1$ but cannot account for those with $\gamma \gg
    1$.\label{figs:rc_allgamma}}
\end{figure}

We address the latter issue in this {\it Letter}, by exploring
numerically the closed orbits of gaseous particles within the potential of an
idealized triaxial CDM halo.  We embed a massless, isothermal gaseous disk
within a cuspy dark matter halo and evolve it until equilibrium is
reached. 

We focus our analysis on the {\it shape} of the rotation curves inferred for the
disk from simulated long-slit observations of the velocity field; on 
deviations from the spherically-averaged circular velocity curve; and on
the possibility that such deviations might account for the discrepancy between
LSB rotation curves and NFW profile fits to CDM halos.

\section{LSB rotation curves}
\label{sec:lsbrc} 

Figure~\ref{figs:rc_allgamma} illustrates the disagreement alluded to
above. This figure shows the H$\alpha$ rotation curves of two LSB galaxies
(points with error bars) selected from the sample of \citet[][B02]{DEBLOK02} and
\citet[][S03]{SWATERS03}. The data points have been fitted using a simple formula,
$V_{\rm rot}(r)=V_0 (1+(r/r_t)^{-\gamma})^{-1/\gamma}$ \citep{COURTEAU97}. Here
$V_0$ and $r_t$ are dimensional scaling parameters, whereas $\gamma$ is a
dimensionless parameter that characterizes the {\it shape} of the rotation
curve. This three-parameter formula provides excellent fits to both LSBs, as
illustrated by the quality of the (solid line) fits shown in
Figure~\ref{figs:rc_allgamma}.

In order to emphasize discrepancies in shape, the rotation curves in
Figure~\ref{figs:rc_allgamma} have been scaled to the radius, $r_{0.3}$, and
velocity, $V_{0.3}$, where the logarithmic slope of the curve is $d\log V_{\rm
rot}/d\log r=0.3$.  The two galaxies shown in
Figure~\ref{figs:rc_allgamma} have different values of $\gamma$, and have been
chosen to illustrate the extreme cases in the B02 and S03
datasets. Roughly one third of their LSBs have $\gamma \lsim 1$, having
rotation curves with shapes similar to UGC 5721 ($\gamma=0.6$).

The dashed line in Figure~\ref{figs:rc_allgamma} shows the $V_c$ profile of an
NFW halo, which is fixed in these scaled units. Galaxies with $\gamma \lsim 1$
are consistent with NFW, whereas those with $\gamma \gg 1$ are clearly
inconsistent.  Figure~\ref{figs:rc_allgamma} also shows the spherically-averaged
$V_c$ profiles of all galaxy-sized halos presented in \cite{HAYASHI04} and
\cite{NAVARRO04}
, scaled to $r_{0.3}$ and $V_{0.3}$.
We find that the shapes of the dark halo $V_c$ curves are in fact quite similar
to NFW.  In terms of the $\gamma$ parameter, most halos (about $95\%$) have
$\gamma \lsim 1$, which implies that LSBs with $\gamma \gg 1$ are quite
difficult to reconcile with the $V_c$ profiles of simulated CDM halos.

On further examination, however, \cite{HAYASHI04} note that most rotation curves
that have best fit values of $\gamma > 1$ also have acceptable fits with $\gamma
\leq 1$.  As a result, only a small minority of LSBs (about $10\%$) are robustly
inconsistent with CDM halo $V_c$ profiles.  Most of these curves are
characterized by a linear rise in velocity with radius and have best fit values
of $\gamma \gsim 5$.  One such example, UGC 5750, is shown in
Figure~\ref{figs:rc_allgamma}, along with a dot-dashed line that illustrates the
$V \propto r$ dependence expected in the presence of a constant density core.

Does this result rule out the presence of a cusp in the dark matter density
profile in such galaxies? As noted in \S~\ref{sec:intro}, before concluding so
one must take into account possible systematic differences between rotation speed
and circular velocity in gaseous disks embedded within realistic, triaxial
halos. This is a complex issue that involves a number of parameters, such as the
degree of triaxiality, the role of the disk's self-gravity, size, and
orientation, as well as the possibility of transient deviations from
equilibrium.

To keep matters simple, we have decided to address this issue by evolving a
massless gaseous disk at the centre of a fixed triaxial halo. We use the
N-body/hydrodynamical code {\tt GASOLINE}, developed by J.Wadsley, J.Stadel, and
T.Quinn \citep{WADSLEY04}. {\tt GASOLINE} combines a tree-based Poisson solver
for gravitational interactions with the Smooth Particle Hydrodynamics (SPH)
technique.  The dark matter halo is modelled with a particle realization of an
NFW mass profile; the mass, virial radius and concentration of the halo are
normalized to $M_{200}= 10^{11}~M_{\odot}/h$, $r_{200}=75~\kpch$ and $c=12$,
respectively.  In order to reduce computational expense, the halo is truncated
at an outermost radius, $r_{\rm trunc} =37.5~\kpch$ and the number of particles
within this radius is set to $1.7 \times 10^5$.  The halo is made triaxial by
multiplying the x-, y-, and z-coordinates of the halo particles by factors of
1.36, 0.82, and 0.68, respectively, so that the spherically-averaged mass
profile remains unchanged, but the axis ratios of the halo are
$0.5\colon0.6\colon1.0$.  We note that such halo shapes (roughly prolate with
elongation 2:1) are among the most common ones found in cosmological
simulations \citep[see, e.g.,][]{JING02}. The gravitational potential is derived
directly from the dark matter particle positions and is assumed to remain
constant in time.

\begin{figure}[t]
\epsscale{1.05}
\plotone{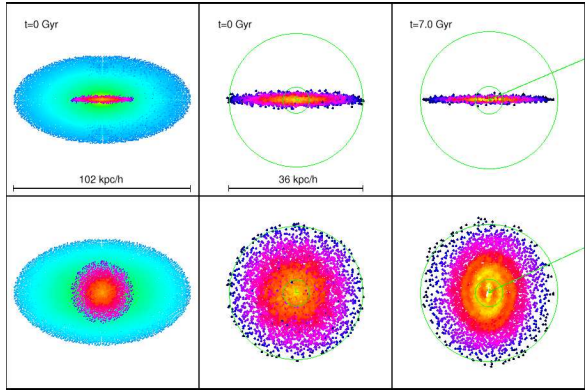}
\caption[Evolution of a massless gaseous disk in an idealized CDM halo]
{Evolution of a massless gaseous disk in the 3D potential of an idealized CDM
halo. Upper (lower) panel shows edge-on (face-on) projections of the disk.  Left
panels shows halo and disk particles at the initial time, color-coded by local
dark matter and gas density, respectively.  Middle and right panels show disk
particles only at initial time and after 7.0 Gyr, respectively.  Inner and outer
circles have radii equal to the exponential scale length of the disk, $r_d$, and
the outermost radius of the disk, $r_{\rm outer}$, at $t=0$, respectively.  The
straight lines in the right panels show projections of the line-of-sight used to
generate the rotation curve in the lower right panel of
Figure~\ref{figs:rotcur}.  Triaxiality in the halo mass distribution leads to
significant evolution in the structure of the disk and to strong deviations from
circular symmetry.\label{figs:evdisk}}
\end{figure}

An exponential disk consisting of $10^4$ massless gas particles is placed at the
centre of the halo, in the plane of the major and intermediate axes of the halo
(see Figure~\ref{figs:evdisk}).  The exponential scale length of the disk is
$r_d = 3.6~\kpch$, and its outermost radius is $r_{\rm outer} = 18.0~\kpch$.
The vertical distribution of the disk material is given by $\rho(z) \propto {\rm
sech}^2(z/z_0)$, with $z_0 = 0.6~\kpch$, and the disk is truncated at $z_{\rm
max} = 3.0~\kpch$. The gas is given a temperature $T=100~K$, well below the
virial temperature of the halo, $T_{\rm vir} \simeq 2
\times 10^5~{\rm K}$, and is assumed to remain isothermal during the evolution.
Pressure forces are thus unimportant for the dynamics of the gas, and the
hydrodynamical treatment simply forces the gas to provide a massless tracer of
the closed orbits within the halo potential.  

\setcounter{footnote}{0}

\begin{figure}[t]
\epsscale{1.05}
\plotone{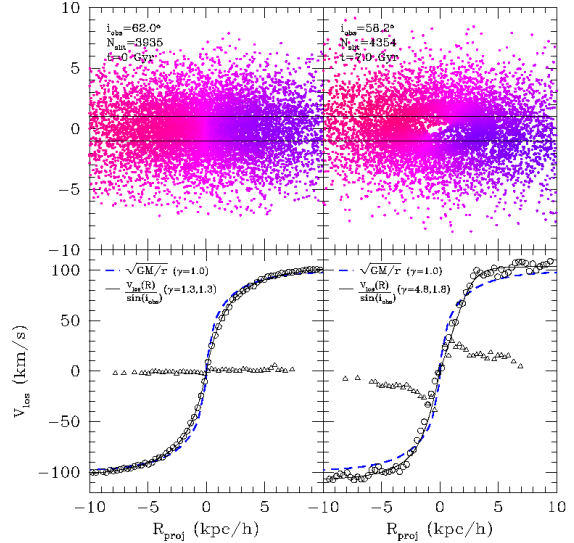}
\caption[Rotation curve of simulated disk as inferred from simulated long-slit
  radial velocity data] {{\it Upper panels:} Projected positions of the disk
  particles after the disk has been inclined by $67^\circ$ relative to the
  initial plane of the disk.  Disk particles are color-coded by line-of-sight
  velocity.  Solid horizontal lines indicate the position of a $2~\kpch$ wide
  slit oriented along the photometric major axis of the projected disk. {\it
  Lower panels:} Rotation curve as inferred from simulated long-slit radial
  velocity data with slit placed across the major axis (open circles) and minor
  axis (open triangles) of the disk.  The major axis rotation curve agrees well
  initially ($t=0$) with the $V_c$ profile of the halo (dotted curves) but
  significant deviations and misalignment of the kinematic and photometric major
  axes result from the evolution of the disk in the triaxial potential of the
  halo.  
\label{figs:rotcur} }
\end{figure}

The disk particles are initially given tangential velocities consistent with the
spherically-averaged circular velocity profile of the triaxial dark matter halo.
As a result the gaseous disk is not initially in equilibrium and it evolves
rapidly to a configuration characterized by departures from circular symmetry
(the disk becomes elliptical, see Figure~\ref{figs:evdisk}). The effect of such
departures from circular symmetry on rotation curves derived from long-slit
spectra is complex, as illustrated in Figure~\ref{figs:rotcur}. This figure
compares the rotation speed of the gas, as inferred from line-of-sight
velocities measured on a slit placed along the photometric major axis of the
disk determined from the projected particle distribution. The disk is inclined
by $67^\circ$, and velocities are corrected by the sine of the inclination
angle, $i_{\rm obs}$, as derived from the aspect ratio of the isodensity
contours of the gas.\footnote{This is inferred to be $\simeq 58^\circ$ because
the disk is not circularly symmetric.}

The left panels in Figure~\ref{figs:rotcur} show that initially, when the disk
is circularly symmetric by construction, the disk inclination is recovered
reasonably accurately and the rotation speed inferred from line-of-sight
velocities (open circles) agree well with the spherically-averaged circular
velocity (dashed line).  Fitting the halo $V_c $ profile out to $r_{\rm outer} =
18~\kpch$ using the $(r_t,\gamma, V_0)$ formula results in best fit values of
$\gamma = 1$, whereas the initial disk rotation curve is best fit by
$\gamma=1.3$. 

At later times the evolution of the disk leads to poorer estimates of the
inclination as well as to significant deviations between inferred rotation speed
and circular velocity.  The {\it shape} of the rotation curve, in particular, is
affected, as shown in the right-hand panels in Figure~\ref{figs:rotcur}. On
some projections, rotation curves appear to rise and turn abruptly, and they
would be (erroneously) taken to imply the existence of a constant-density core
in simple models that assume spherical symmetry. Fits to the disk rotation
curves using the ($r_t,\gamma,V_0$) formula introduced above often have $\gamma
\gg 1$, consistent with galaxies where NFW profiles provide a particularly poor
fit to the rotation curve data.  The open triangles in the lower panels of
Figure~\ref{figs:rotcur} show the rotation curves obtained with the slit oriented
along the minor axis of the disk.  The misalignment between the photometric and
kinematic axes of the disk  after 7.0 Gyr is clearly evident.

We have calculated the distribution of $\gamma$ values for rotation curves
obtained by observing the simulated disk at time $t=7.0$ Gyr from 1000 random
lines-of-sight with inclination angles limited to $30^\circ < i <
70^\circ$.  A large fraction of rotation curves have values of $\gamma$
significantly higher than the initial value of $\gamma =1.3$.  To be precise,
approximately $45\%, 25\%,$ and $7\%$, have $\gamma >2, >3,$ and $>4$,
respectively.  Furthermore, minor axis rotation is seen with a peak amplitude
exceeding 20\% of the major axis amplitude in 70\% of simulated rotation curves
with $\gamma > 3$.
 
\begin{figure}[t]
\epsscale{1.05}
\plotone{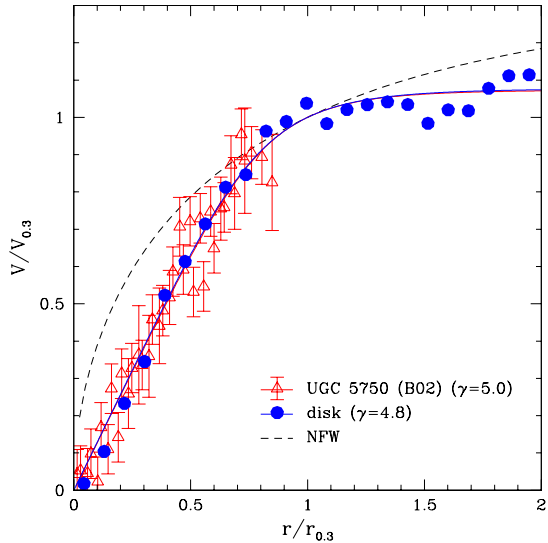}
\caption
[Rotation curves of simulated disk compared with those of LSB galaxies]
{Systematic deviations from circular velocity induced by the
triaxiality of the halo lead to rotation curve shapes in better agreement with
LSBs with high $\gamma$. Open circles correspond to rotation curves inferred
from simulated long-slit data for the disk shown in Figure~\ref{figs:rotcur}.
\label{figs:rc_allgamma_sim}}
\end{figure}

In Figure~\ref{figs:rc_allgamma_sim}, we compare the high-$\gamma$ LSB rotation
curve shown in Figure~\ref{figs:rc_allgamma} with one obtained from the
projection of the disk shown in Figure~\ref{figs:rotcur}. The agreement between
the simulated and observed rotation curves is excellent, suggesting that
deviations from spherical symmetry in CDM halos might reconcile disk rotation
curves that appear to favour the presence of constant density cores with cusps
in the dark matter density profiles.

\section{Discussion}
\label{sec:disc}

Given that we are able to match discrepant rotation curve shapes with a disk
embedded in a cuspy triaxial halo, the outlook for reconciling dark matter cusps
with LSB rotation curves is rather encouraging. However, it would be premature
to argue that the problem has been fully solved. After all, given the number of
extra ``free'' parameters introduced by relaxing the assumption of spherical
symmetry, it is perhaps not surprising that one is able to improve the agreement
with LSB rotation curves.

It is therefore important to build a more compelling case for this
interpretation of LSB rotation curves, so as to render it falsifiable. Are there
any corroborating traits that may be used to confirm or exclude the hypothesis
that halos surrounding LSBs are indeed triaxial? In particular, we would like to
understand better the particular combination of perspective and triaxiality that
results in rotation curves with values of $\gamma >1$. How can one best verify
the triaxial-halo interpretation in two-dimensional velocity maps?  Identifying
a clean and unambiguous indication of triaxiality, such as the unusual minor
axis kinematics shown in Figure~\ref{figs:rotcur}, will be as important as the
success of aspherical halos in reproducing the rich variety of shapes of LSB
rotation curves.  Only if this is accomplished shall we be able to conclude that
LSB rotation curves do not preclude the presence of dark matter density cusps,
thereby freeing the CDM paradigm of one vexing challenge on small scales.

\acknowledgements This work has been supported by various grants to JFN from
NSERC, CFI, and by fellowships from the Alexander von Humboldt Foundation.

\bibliography{stan}

\end{document}